\documentstyle[12pt,cite,a4,twoside]{article}

\let\Huge=\large
\let\large=\normalsize

\newcommand{\imag}{{\rm i}}
\newcommand{\fm}{\,{\rm fm}}
\newcommand{\GeV}{\,{\rm GeV}}
\newcommand{\MeV}{\,{\rm MeV}}
\newcommand{\mb}{\,{\rm mb}}
\newcommand{\de}{{\rm d}}

\renewcommand{\vec}[1]{\mathord{\bf #1}}
\newcommand{\llongrightarrow}
{\relbar\joinrel\relbar\joinrel\longrightarrow}
\newcommand{\lsim}{
 \mathrel{\setbox0=\hbox{$<$}\raise0.6ex\copy0\kern-\wd0
 \lower0.65ex\hbox{$\sim$}}}

\begin{document}

\begin{titlepage}
\vspace*{-2cm}
\begin{flushright}
\bf
ADP-95-10/T173\\
TUM/T39-95-1
\end{flushright}
\begin{center}
  {\Huge \bf Phenomenology of Nuclear
    Shadowing in Deep-Inelastic
    Scattering$^{*}$}

  \vfill

  {\large G. Piller$^{a,c}$, W.
    Ratzka$^{b}$ and W. Weise$^{c}$ }
  \bigskip

  $^{a)}$Department of Physics and
  Mathematical Physics, University of
  Adelaide,\\ S.A. 5005, Australia\\
  $^{b)}$Institut f\"ur Theoretische
  Physik, Universit\"at Regensburg,\\
  D-93040 Regensburg, Germany\\
  $^{c)}$
  Physik Department, Technische
  Universit\"at M\"unchen,\\ D-85747
  Garching, Germany

  \vfill

  {\bf Abstract}
\end{center}

We investigate shadowing effects in
deep-inelastic scattering from nuclei at
small values $x<0.1$ of the Bjorken
variable.  Unifying aspects of generalized
vector meson dominance and color
transparency we first develop a model for
deep-inelastic scattering from free
nucleons at small $x$.
In application
to nuclear
targets we find that the coherent
interaction of quark-antiquark
fluctuations with
nucleons in a nucleus
leads to the observed shadowing at
$x<0.1$.  We compare our results with most
of the recent data for a large variety of
nuclei
and examine in particular
the $Q^2$ dependence of the shadowing
effect.
While
the coherent interaction of
low mass vector mesons
causes
a major part of the shadowing observed in
the $Q^2$ range of current experiments,
the coherent scattering of continuum
quark-antiquark pairs is also important
and guarantees a very weak overall $Q^2$
dependence of the effect.  We also discuss
shadowing in deuterium and its
implications for the quark flavor
structure of nucleons.  Finally we comment
on shadowing effects in high-energy
photon-nucleus reactions with real
photons.

\vfill
\begin{center}
\it To be published in Z. Phys. A.
\end{center}

\vfill
\noindent $^{*}$) Work supported in part
by grants from BMFT, DFG and ARC.

\end{titlepage}

\section{Introduction}

In recent years
numerous
experiments
have
been dedicated to high
precision measurements of deep-inelastic
lepton scattering from nuclei.
Experiments at CERN
\cite{NMC91,NMC92,NMC94,Paic94} and
Fermilab
\cite{E66592prl,E66592plb,Carroll93,Spentz94}
focus especially on the region of small
values of the Bjorken variable $x =
Q^2/2M\nu$, where $Q^2 = - q^2$ is the
squared four-momentum transfer, $\nu$ the
energy transfer and $M$ the nucleon mass.
The data, taken over a wide kinematic
range from $10^{-5} <x< 0.65$ and $0.01
\GeV^2 < Q^2 < 100\, \GeV^2$, show a
systematic reduction of the nuclear
structure function $F_2^A(x,Q^2)$ with
respect to $A$ times the free nucleon
structure function $F_2^N(x,Q^2)$ at
$x<0.1$.

This so-called shadowing effect
has prompted a fair amount of theoretical
actvity
(for recent reviews see
e.g.~\cite{BadKwi92rmp}).
Some of the existing work focuses on
an infinite momentum frame description of
the scattering process
(see e.g.~\cite{MueQiu86}).  The driving
mechanism in these models is given by
quark and gluon annihilation processes
at high parton densities, which are
described using perturbative techniques.
Although these methods allow one to
address the $Q^2$ dependence of the
shadowing effect, its $x$ dependence is
not accessible to perturbation theory and
therefore subject to parametrization.

Another class of models considers the
deep-inelastic scattering process in the
laboratory frame where the target is at
rest
\cite{SakSch72,Shaw89,BaSpYe78,
FraStr89,
BrodLu90,NikZak91zpc,BadKwi92npb,
MelTho93,
KuPiWe94,
PilWei90}.  In this frame the interaction
at small values of $x$ proceeds
via hadronic
components present in the wave function of
the exchanged virtual photon.  The
coherence length of these hadronic
fluctuations is typically of order $1/Mx$
and
exceeds, for $x<0.1$,
the average
nucleon-nucleon distance in nuclei.
Hence for small $x$
the hadronic fluctuations will interact
coherently with several nucleons inside
the target nucleus.
Shadowing is then caused by destructive
interference of multiple scattering
amplitudes which describe the passage of
these fluctuations through the nucleus.

In this paper we present a laboratory
frame description of deep-inelastic
scattering at small $x$, based on ideas
which unify the generalized vector meson
dominance picture \cite{DonSha78} with the
concept of color transparency (for recent
reviews see
\cite{Nikola92}).  At small momentum transfers, $Q^2<1
\GeV^2$, the hadronic components of the
absorbed virtual photon are formed by
strongly correlated quark-antiquark pairs,
most prominently by the low mass vector
mesons $\rho$, $\omega$ and $\phi$.
At larger $Q^2$
quark-antiquark pairs from the so-called
$q\bar q$-continuum
become increasingly important.
We include
both
strongly
correlated and
continuum
$q\bar q$-fluctuations in terms of the measured
photon spectral function.  While some
empirical information is available about
the interaction of low mass vector mesons
with nucleons and nuclei, the interaction
properties of
continuum
quark-antiquark pairs are
scarcely known. To fill this gap we use
color transparency as a guiding principle,
i.e.\ the cross section of
color singlet quark-antiquark pairs is
assumed to be proportional to their
transverse size.  With these ingredients
we obtain a good description of both
nucleon and nuclear structure functions,
$F_2^N(x,Q^2)$ and $F_2^A(x,Q^2)$, at
small $x$.

Although some of the ideas mentioned above
are common to several recent models of
deep-inelastic scattering at small $x$,
little effort has been directed towards a
quantitative comparison with the now
available large amount of experimental
data.  We confront our model with most of
the recent data.  In particular we examine
the $Q^2$ dependence of the shadowing
effect -- an intensely discussed issue.
We find that while a major part of the
shadowing seen in current experiments is
caused by the coherent interaction of low
mass vector mesons, coherent scattering of
continuum
$q\bar q$ pairs is also
important and guarantees a weak $Q^2$
dependence of the shadowing effect.

We will also briefly discuss shadowing
effects in high-energy photon-nucleus
reactions with real photons (i.e.\ in the
limit $Q^2\rightarrow 0$).

The plan of this paper is as follows: In
Section 2 we introduce the space-time
picture of deep-inelastic scattering in
the laboratory frame. First we develop a
model for deep-inelastic scattering from
free nucleons in Section 3. Its extension
to nuclear targets is described in Section
4. We discuss shadowing in nuclei with
intermediate and large masses as well as
its implications for a deuterium target.
In Section 5 we apply our model to high
energy photon-nucleus scattering.
Finally, Section 6 contains a summary and
conclusions.

\section{Lab frame picture of deep-inelastic
  scattering}\label{sec:labframe}
It is common to discuss deep-inelastic
lepton scattering on free nucleons in a
frame where the target moves with a large
momentum, $|\vec{p}|\to\infty$.  In this
{\em infinite momentum frame\/} the parton
model allows the interpretation of
measured structure functions as momentum
distributions of quarks and antiquarks in
the target.  There is, however, no
reliable approach for dealing with nuclear
systems in this frame.  Consequently the
preferable frame of reference for an
investigation of nuclear effects in
deep-inelastic scattering is rather the
{\em laboratory system\/} in which the
target is at rest.  Well established
knowledge about the structure and geometry
of nuclear targets can then be used.

Consider therefore a description of
deep-inelastic lepton scattering from
nucleons or nuclei in the laboratory
frame.  Here the basic photon-nucleon
interaction process involves the time
orderings
shown in
Figs.~\ref{fig:timeorder}(a)
and~\ref{fig:timeorder}(b): the photon
either hits a quark (or antiquark) in the
target which picks up the large energy and
momentum transfer, or the photon converts
into a quark-antiquark pair that
subsequently interacts with the target.

For small $x$ the pair production process
(b) dominates the scattering amplitude, as
can be shown in time-ordered perturbation
theory~\cite{BaSpYe78}: The amplitudes
${\cal A}_a$ and ${\cal A}_b$ of processes
(a) and (b) are roughly proportional to
the inverse of their corresponding energy
denominators $\Delta E_a$ and $\Delta
E_b$.  For large energy transfers $\nu \gg
M$ these are:
\begin{eqnarray}
  \Delta E_a &=& E_a(t_2)-E_a(t_1) \approx
  -\left<p_q^2\right>^{1/2}
  +\frac{\frac{2}{3}\left<p_q^2\right>+Q^2}{2\nu},
\label{eq:deltaEa}\\
\Delta E_b &=& E_b(t_2)-E_b(t_1)
\approx\frac{\mu^2+Q^2}{2\nu}\;,
\label{eq:deltaEb}
\end{eqnarray}
where $\left<p_q^2\right>^{1/2}$ is the
average quark momentum in a nucleon and
$\mu$ is the invariant mass of the
quark-antiquark pair.  We then obtain for
the ratio of these amplitudes:
\begin{eqnarray} \label{amplitudes}
\left|\frac{{\cal A}_a}{{\cal A}_b}\right|
\sim
\left|\frac{\Delta E_b}{\Delta E_a}\right|
\approx
 \frac{M x}{\left<p_q^2\right>^{1/2}}
\left(1 + \frac{\mu^2}{Q^2}\right).
\end{eqnarray}
As we will argue later, the main
contribution to process (b) comes from
quark-antiquark pairs with
a
squared mass
$\mu^2 \sim Q^2$.  The ratio in
Eq.~(\ref{amplitudes}) is evidently small
compared to unity for $x\ll 0.1$.  Hence
pair production,
Fig.~\ref{fig:timeorder}(b), will be the
dominant lab frame process in the
small-$x$ region.
For the following discussion of
deep-inelastic scattering at small $x$ we
will therefore consider the dominant
process (b) only (although it should of
course be noted that in principle only the
sum of (a) and (b) is Lorentz and gauge invariant).

What are the implications of this picture
for deep-inelastic scattering from nuclear
targets?  The coherence length $\lambda$
of a photon-induced hadronic fluctuation
with mass $\mu$ is given by the inverse of
the energy denominator~(\ref{eq:deltaEb}):
\begin{equation}
\lambda \sim \frac{1}{\Delta E_b}
=
\frac{2\nu}{\mu^2+Q^2}
\stackrel{\mu^2\sim Q^2}{\llongrightarrow}
\frac {1}{2 x M}.
\label{eq:coherence}
\end{equation}
For $x<0.05$ this coherence length exceeds
the average distance between nucleons in
nuclei, $d\approx1.8\fm$.  Then coherent
multiple scattering on several nucleons in
the target can occur, leading to nuclear
shadowing.

For larger $x$, the coherence length of
the intermediate $q\bar q$ state is small,
$\lambda < d$.  At the same time the
process in Fig.~\ref{fig:timeorder}(a)
becomes prominent, i.e.\ the virtual
photon is absorbed directly by a quark or
antiquark in the target.  In the range of
intermediate and large $x$, say $x>0.2$,
the virtual photon therefore interacts
incoherently with nucleons bound in the
nuclear target.

\section{Deep-inelastic scattering from
  free nucleons at small
  $x$}\label{sec:nucleon}
Before discussing effects in the
scattering from nuclei
we have to develop a
model of
deep-inelastic  scattering from
free nucleons.
The free nucleon structure function
$F_2^N(x,Q^2)$, defined as the average of
the proton and the neutron structure function,
can be written in terms of the virtual
photon-nucleon cross section
$\sigma_{\gamma^* N}$:
\begin{equation}
F_2^N(x,Q^2) = \frac{1-x}{1+\frac {Q^2}{\nu^2}}
\,\frac{Q^2}{4\pi^2 \alpha_{{\rm em}}}
\sigma_{\gamma^*N}.
\label{eq:sigmagammaN1}
\end{equation}
In the limit $\nu^2\gg Q^2$ and $x<0.1$ that we are
concerned with, this simplifies to
\begin{equation}
F_2^N(x,Q^2) = \frac{Q^2}{4\pi^2 \alpha_{{\rm em}}}
\sigma_{\gamma^*N}
\label{eq:sigmagammaN}
\end{equation}
(we use $\alpha_{{\rm em}} \equiv e^2/4\pi = 1/137$).
We will consider only
contributions from transversely polarized
photons, as they constitute the dominant part
of the cross section:
$\sigma_{\gamma^* N}\approx\sigma^T_{\gamma^* N}$
(see \cite{DaBaea94} for an experimental analysis
of $\sigma_L/\sigma_T$).

As discussed above, the virtual photon
interacts with the nucleon by first converting into
a $q\bar q$ pair which then propagates,
forming a hadronic intermediate
state that interacts strongly with the
nucleon.
This is expressed in the following
spectral  ansatz for the structure
function
\cite{SakSch72,FraStr89,Gribov70,BjoKog73,PilWei90}
valid at~${x<0.1}$:
\begin{eqnarray}
F_2^N(x,Q^2)=
 \frac{Q^2}{\pi}
 \int_{4m_{\pi}^2}^\infty\,\de\mu^2
 \frac  {\mu^2 \,\Pi(\mu^2)}
        {\left(\mu^2+Q^2\right)^2}\;
 \sigma_{hN}(\mu^2).
\label{eq:xVDM}
\end{eqnarray}
Here $\Pi(\mu^2)$ is the spectrum of
hadronic fluctuations with mass $\mu$ which
is related to the measured cross section
for $e^+e^-\to {\rm hadrons}$ by
\begin{equation}
\Pi(s) = \frac{1}{12 \pi^2}
         \frac{\sigma_{e^+e^-\rightarrow {\rm hadrons}}(s)}
          {\sigma_{e^+e^-\rightarrow \mu^+\mu^-}(s)}.
\end{equation}
Note that the effective hadron-nucleon cross
section $\sigma_{hN}(\mu^2)$ in
Eq.~(\ref{eq:xVDM})
is an average including all contributions
with a given invariant mass $\mu$.
The factor $(\mu^2+Q^2)^{-2}$
in Eq.~(\ref{eq:xVDM}) comes from the
propagators of the hadronic intermediate states.
One should of course note that Eq.~(\ref{eq:xVDM})
cannot be expected to follow directly from
perturbative QCD. In particular, the
effective cross section
$\sigma_{hN}(\mu^2)$ incorporates
non-perturbative physics characteristic of
the small-$x$ region. However, as will be
shown,  Eq.~(\ref{eq:xVDM}) does have the
proper logarithmic behavior at large $Q^2$.

The structure function $F_2^N(x,Q^2)$ in
Eq.~(\ref{eq:xVDM}) is dominated by
contributions from
intermediate states
with an invariant mass $\mu^2\sim Q^2$. As a
consequence for small momentum transfer, $Q^2
< 1\GeV^2$, the low mass vector
mesons $\rho,\;\omega$ and $\phi$ are of
major importance. They represent strongly
correlated quark-antiquark pairs which contribute
the term
\begin{equation} \label{eq:Pivmd}
\Pi^{{\rm VMD}}(q^2)
= \sum_{V=\rho,\omega,\phi}
\left(
        \frac {m_V}{g_V}
\right)^2
\delta(q^2-m_V^2)
\end{equation}
to the photon spectral function.
Here $m_V$ are the vector meson masses and
$g_V^{-1}$ the corresponding $\gamma V$
coupling constants
(see Table~\ref{tab:vmesons}).
Eq.~(\ref{eq:Pivmd}) represents the traditional Vector Meson Dominance
(VMD) model.
For large $Q^2$ the heavy vector mesons $J/\psi$ and $\psi'$
also contribute, and we take them into account as well.
Altogether, vector mesons give a
contribution to the nucleon structure
function of the form
\begin{equation}
 F_2^{N,{\rm VMD}}(x,Q^2)
 =
   \frac{Q^2}{\pi}
     \sum_{V}
     \left(
         \frac {m_V^2}{g_V}
     \right) ^2
     \left(\frac{1}{m_V^2+Q^2}\right)^2\,
     \sigma_{VN}.  \label{F2VMD}
\end{equation}

\noindent Here $\sigma_{VN}$ are the vector meson-nucleon
cross sections.
They can be determined in real and virtual
photoproduction experiments (see
Table~\ref{tab:vmesons}). It should
be mentioned that their exact value may
in principle depend  on the kinematics
of the experiment (see e.g.~\cite{KoNeNi93}).

For small values of $x$ and $Q^2$
($x<0.1$ and $Q^2<1\GeV^2$) the
interactions of the low mass vector mesons
dominate  the nucleon structure function
$F_2^N$ and lead to the scale breaking
behavior $F_2^N(x,Q^2)\sim Q^2$
for
$Q^2 \rightarrow 0$.

For larger values of the momentum transfer, i.e.
$Q^2>m_{\phi}^2\approx 1\GeV^2$,
the nucleon structure function $F_2^N$ is
governed by the interaction of
quark-antiquark pairs with
mass $\mu^2\sim Q^2>1\GeV^2$. Apart from
the narrow charmonium
and upsilon resonances, these quark pairs
form the so called $q\bar q$ continuum.
In the annihilation of $e^+e^-$ into hadrons
they are responsible for the
approximately constant
behavior of the cross section ratio
at large timelike momenta,
$
\sigma_{e^+e^-\to {\rm hadrons}}
/\sigma_{e^+e^-\to \mu^+\mu^-}
\approx3{\sum_f e_f^2}$,
where we sum over the fractional charges
$e_f$ of all quark flavors
which are energetically
accessible.

To calculate the contribution of the
continuum
quark-antiquark fluctuations
to the nucleon structure function we need
to know their effective interaction cross section.
Since the $q\bar q$ fluctuations of the  photon
are color singlets,
we assume their cross sections to scale
with their transverse size $\rho$ (i.e.\
their size in a plane perpendicular to
their momentum)
as $\sigma\sim\rho^2$~\cite{NikZak91zpc}.
Investigating  the geometry of the
dissociation of a photon into an
uncorrelated $q\bar q$ pair more closely
(see e.g.~\cite{DavNik78}) one obtains the following
approximate expression for $\rho^2$:
\begin{equation}     \label{eq:rhosquare}
\rho^2\approx
\frac{1}{\alpha(1-\alpha)}\,
\frac{4\mu^2}{(\mu^2+Q^2)^2}.
\end{equation}
Here $\alpha$ is the fraction of the light-cone
momentum carried by the quark:
if $k_q$ is the quark momentum and
$q=(q_0,\vec{0}_{\perp},q_3)$ the photon momentum,
one has $\alpha = (k_{q0} + k_{q3})/(q_0 +
q_3)$. (Correspondingly, $1-\alpha$ is the
light-cone momentum fraction carried by the
antiquark.)  Of course Eq.~(\ref{eq:rhosquare})
is a reasonable estimate for the size
of the $q\bar q$ fluctuation only as long as the
distance $\rho$ of the pair is smaller than
a typical confinement scale
of about $1\,\fm$.  If the
distance $\rho$ increases further,
strong interactions between the quark and
antiquark will limit the transverse size of
the $q\bar q$ fluctuation, thus
leading to a saturation of $\rho$.
Having this in mind, we choose for the effective cross
section of continuum quark-antiquark pairs:
\begin{equation} \label{sigcon}
\sigma_{hN}(\mu^2,\alpha)= K\cdot \rho^2 =
K\cdot \min
\left\{\begin{array}{l}
    R_c^2 \\
    \displaystyle\frac{1}{\alpha(1-\alpha)}\;
    \frac{4\mu^2}{(\mu^2+Q^2)^2},
\end{array}\right.
\end{equation}
with a constant $K$ to be determined.
Here we have introduced a maximum radius $R_c$
which should be in the range of the confinement
scale.
As it is clear from our discussion above,
the cross section $\sigma_{hN}(\mu^2,\alpha)$
depends not only on the invariant mass
$\mu$ of the $q\bar q$ pair, but
also on $\alpha$, i.e.\ on the way the
photon momentum is split between the quark
and antiquark. From Eq.~(\ref{sigcon}) we
observe that the average interaction cross
section of $q\bar q$ pairs with mass $\mu$
is
${\sigma_{hN}(\mu^2) = \int_0^1 \de\alpha
\,\sigma(\mu^2,\alpha) \sim 1/\mu^2}$
(ignoring terms $\sim \log\mu^2$), which
is the behavior necessary for
scaling~\cite{BjoKog73,NikZak91zpc,FraStr89,PilWei90}.

If we now take into account both the vector mesons and
the
quark-antiquark continuum,
we obtain from
Eqs.~(\ref{eq:xVDM},~\ref{F2VMD},~\ref{sigcon})
the following expression for the nucleon structure
function:
\begin{eqnarray}  \label{F2N}
&& F_2^{N}(x,Q^2) =
   \frac{Q^2}{\pi}
     \sum_{V=\rho,\dots}
     \left(\frac {m_V^2}{g_V}\right)^2
     \left(\frac{1}{m_V^2+Q^2}\right)^2
     \,\sigma_{VN} \nonumber                 \\
&&\qquad+ \frac{Q^2}{\pi}
 \int_{{\mu_0}^2}^\infty\,\de\mu^2
 \int_0^1\, \de\alpha
 \frac  {\mu^2 \,\Pi^{{\rm cont.}}(\mu^2)}
        {\left(\mu^2+Q^2\right)^2}\;
 \sigma_{hN}(\mu^2, \alpha),
\end{eqnarray}
valid at small Bjorken $x$.
Here $\Pi^{{\rm cont.}} = \Pi - \Pi^{{\rm VMD}}$
is the continuum part of the photon spectral function which
starts  at $\mu_0^2 \sim m_{\phi}^2$.
While the vector meson part vanishes as $1/Q^2$
for large $Q^2$, the $q\bar q$ continuum
contribution to the structure function
displays logarithmic
scaling behavior:
\begin{equation}
F_2^N(x,Q^2)
\sim\ln\left(R_c^2 Q^2\right)\quad
\mbox{for} \; Q^2 \gg 1\GeV^2.
\end{equation}
We now compare our result for the free
nucleon structure function $F_2^N$ with
recent  data of the New Muon
Collaboration~\cite{NMC92}.  We include in
Eq.~(\ref{F2N}) all vector mesons
$\rho,\,\omega,\,\phi,\,J/\psi$ and
$\psi'$.  Their masses, coupling constants
and cross sections are summarized in
Table~\ref{tab:vmesons}.

The effective cross section $\sigma_{VN}$ (i.e.\ the
forward scattering amplitude)
may depend on the momentum and energy
transfer variables $Q^2$ and $\nu$.
The relevant range in $Q^2$ is, however,
restricted by the fact that the vector
mesons contribute mainly in the
region $Q^2\approx m_V^2$.
On the other hand
experimental constraints
\cite{NMC91,NMC92} put bounds on
the accessible values of $\nu$.
We therefore chose the cross sections to
be approximately constant, setting
$\sigma_{\rho N}=22\mb$
and fixing the other cross sections to
scale like $\sigma_{VN}\sim 1/m_V^2$.

The constant $K$ in
Eq.~(\ref{sigcon}) is fixed at $K=1.7$ together with
$R_c=1.3\fm$. This corresponds to a maximum
value of about $29\mb$ for the effective cross
section of a $q\bar q$-pair interacting
with a nucleon.

{}From Fig.~\ref{fig:fitnucleon} one can see that
our model reproduces the measured nucleon
structure function at small $x$ quite well.
We want to emphasize again the
importance of vector mesons at small values
of $Q^2$. In detail we find that at $Q^2 =
1\GeV^2$ almost half of $F_2^N$
at $x=0.01$ is due to vector mesons. At
$Q^2= 10\GeV^2$ they still contribute
around
$15\%$.

\section{Deep-inelastic scattering from nuclei at
         small $x$}

Just like  the scattering from free nucleons,
scattering from nuclear targets at small values
of $x$ proceeds via the interaction of hadronic
components present in the spectral function of the
exchanged photon.
For $x<0.1$ the nuclear structure function $F_2^A$
can therefore be written in a way
analogous to $F_2^N$ in Eq.~(\ref{F2N}):
\begin{eqnarray}
 F_2^{A}(x,Q^2) =
     \frac{Q^2}{\pi} \sum_{V=\rho,\dots}
     \left(\frac {m_V^2}{g_V}\right)^2
     \left(\frac{1}{m_V^2+Q^2}\right)^2
     \,\sigma_{VA} &&\nonumber \\
 + \frac{Q^2}{\pi}
 \int_{{\mu_0}^2}^\infty\,\de\mu^2
 \int_0^1\, \de\alpha \,
 \frac  {\mu^2 \Pi^{\rm cont.}(\mu^2)}
        {\left(\mu^2+Q^2\right)^2}\;
 \sigma_{hA}(\mu^2, \alpha). && \label{eq:F2A}
\end{eqnarray}
We have just replaced the hadron-nucleon cross
sections $\sigma_{hN}$ in Eq.~(\ref{F2N}) by the
corresponding hadron-nucleus cross sections
$\sigma_{hA}$.

As mentioned in
Section~\ref{sec:labframe}, for
$x<0.05$
the coherence length $\lambda =
2\nu/(Q^2+\mu^2)$ of the interacting
hadronic fluctuation exceeds the average
internucleon distance in nuclei.
Consequently the intermediate hadronic
system can scatter coherently from several
nucleons in the target.  Interference
between the multiple scattering amplitudes
causes a reduction of the hadron-nucleus
cross sections compared to the na{\"\i}ve
result of just $A$ times the respective
hadron-nucleon cross section and thus
leads to shadowing.

These effects are described by the
Glauber-Gribov multiple scattering formalism
\cite{Bertoc72} which we
will
now summarize briefly.

\subsection{Glauber-Gribov multiple scattering
theory}
\label{sec:glauber}

Let us consider high energy forward scattering
of a hadronic fluctuation $h$ with four-momentum
$q = (\nu,\vec{0}_{\perp},\sqrt{\nu^2 + Q^2})$
and mass $\mu$  on a nucleus. In the laboratory
frame the target momentum
is $P = (A(M-{\cal E}),\vec 0)$,
where $A$ is the nuclear mass number and
${\cal E}$ the binding energy per nucleon.
The scattering amplitude ${\cal A}_{hA}$
for  this process can be written as the
sum
${{\cal A}_{hA} = \sum_{n=1}^A {\cal
A}_h^{(n)}}$
over multiple scattering terms ${\cal
A}_h^{(n)}$, each of which describes
the projectile interacting consecutively with
$n$ nucleons in the target
(see Fig.~\ref{fig:ms1}):
\begin{eqnarray}
{\cal A}_h^{(n)}&=&
\frac{A!}{(A-n)!}
\prod_{i=1}^{n-1}
   \left[
     \int\frac{\de l_{i}}{(2\pi) 2(M-{\cal E})}
   \right]
{V}_h^{(n)}(\nu,\dots q_{iz}\dots)
                                            \nonumber\\
&&\times
\int \de^2b\;
\int_{-\infty}^{\infty} \de z_1
\int_{z_1}^{\infty} \de z_2
\cdots
\int_{z_{n-1}}^{\infty} \de z_n
                                            \nonumber\\
&&\qquad\times
\rho_n(\vec b,z_1\dots z_n)
\prod_{i=1}^{n-1}
\left[
        {\rm e}^{\imag l_i(z_i-z_{i+1})}
\right] . \label{eq:mulscat:zord}
\end{eqnarray}
Here
${V}_h^{(n)}(\nu,\dots q_{iz}\dots)$
describes the
interaction of the hadronic projectile
with $n$ nucleons. For large projectile energies
$\nu$, it is assumed that $V$ depends only
on $\nu$ and
$q_{iz}$,
the
longitudinal momenta transferred
to the interacting nucleons.
Since we consider forward scattering only,
we have
$\sum_{i=1}^n q_{iz} = 0$.
Furthermore
the integration variables $l_i$ are defined as
$l_i=\sum_{j=1}^{i}q_{jz}$,
such that
$\left|\vec{q}\right|-l_i$
is the longitudinal
momentum of the projectile after its
interaction with the $i$th nucleon.
The  nucleon distribution in Eq.~(\ref{eq:mulscat:zord})
is given by the $n$-particle density
\begin{eqnarray}
\rho_n(\vec{b};z_1\dots z_n) =
\prod_{j=n+1}^{A}
\left[
        \int \de^3x_j
\right]
\delta^3 (\vec X_{{\rm cm}})
&& \nonumber\\
\times
\Bigl|
        \Psi\left(
                \vec{b}, z_1;
                \dots;
                \vec{b}, z_n;
                \vec{x}_{n+1}\dots
                \vec{x}_A
        \right)
\Bigr|^2, && \label{eq:npartdens}
\end{eqnarray}
where $\Psi (\dots \vec x_i\dots)$ is
the coordinate-space wave function
of the nucleus.  Its  center of mass
$\vec X_{{\rm cm}} =
\frac{1}{A}\sum_{i=1}^{A}
\vec x_i$  is fixed at the origin.
Since the high energy scattering process
occurs at a fixed impact parameter
$\vec b$, the active nucleons enter $\rho_n$  at
coordinates $\vec x_i = (\vec b,z_i)$
for~${i=1\dots n}$.

As a next step the amplitude ${V}_h^{(n)}$
is expanded in hadronic eigenstates.
Let us denote the complete set of states
that can be reached
after the interaction
with the $i$th nucleon by $\{h_i\}$
and write the corresponding invariant masses
as $m_{h_i}$.
If the conversion from state $h_i$
into state $h_{i+1}$ in the
interaction with the $(i+1)$th nucleon is
described by the transition amplitude
$f_{h_ih_{i+1}}$,
the expression for ${V}_h^{(n)}$ becomes:
\begin{eqnarray}
 \imag {V}_h^{(n)}&=&
\sum_{h_1,\dots,h_{n-1}}
      \imag f_{h h_1}\frac{\imag}{\nu^2-(\left|
\vec{q}\right|-l_1)^2-m_{h_1}^2+\imag\epsilon }
\imag f_{h_1 h_2}
                              \nonumber\\
      &\times& \frac{\imag}{\nu^2-(\left|
      \vec{q}\right|-l_2)^2- m_{h_2}^2+
      \imag\epsilon } \imag f_{h_2 h_3}
       \times\cdots
                              \nonumber\\
       \cdots &\times&
       \frac{\imag}{\nu^2-(\left|\vec{q}\right|
              -l_{n-1})^2-m_{h_{n-1}}^2+ \imag\epsilon }
       \imag f_{h_{n-1} h}.
\label{eq:mulscat:V(n)}
\end{eqnarray}
We can now perform the integration over
the variables $l_i$. We note that the
exponential factors in
Eq.~(\ref{eq:mulscat:zord}) require
that the integration contour be closed
in the lower plane. Picking up the poles,
the longitudinal momentum transfer gets fixed at
\begin{equation}
l_i=\left|\vec{q}\,\right|-\sqrt{\nu^2-m_{h_i}^2}
\approx \frac{Q^2+m_{h_i}^2}{2\nu}.
\end{equation}

For intermediate and heavy nuclei
we may in good approximation
consider only
elastic rescattering of the incoming
hadronic state $h$ from the nucleons
inside the target.
Contributions of inelastically produced
states to multiple scattering were
investigated by Murthy
et~al.~\cite{MuAyGu75} and
Nikolaev~\cite{Nikola86} for high energy
hadron-nucleus scattering processes.
They found such contributions
to be small, though rising logarithmically
with the projectile energy $\nu$.
For example at $\nu\sim 100\GeV$ inelastic
terms account typically for $\sim 5\%$ of
the total hadron-nucleus cross sections
under consideration.

In the so-called ``diagonal approximation'',
i.e.\ neglecting inelastic intermediate states,
the amplitude ${V}_h^{(n)}$ reduces to:
\begin{equation}
\imag {V}^{(n)}_{h\,{\rm diag.}}=
\prod_{i=1}^{n-1}
\left[
  \imag f_{hh}\frac{\imag}{\nu^2-(\left|
\vec{q}\,\right|-l_i)^2-\mu^2+ \imag\epsilon}
\right] \imag f_{hh},
\label{eq:mulscat:V(n)diag}
\end{equation}
In this case the longitudinal momentum transfer
is fixed just at the inverse of the coherence
length $\lambda$
(\ref{eq:coherence})
of the hadronic projectile:
\begin{equation}
l=\left|\vec{q}\,\right|-\sqrt{\nu^2-\mu^2}
\approx \frac{Q^2+\mu^2}{2\nu}= 1/{\lambda}.
\end{equation}

Summing  over all multiple scattering terms
${\cal A}_h^{(n)}$ and neglecting the binding energy
${\cal E}\ll M$ we
find  for the hadron-nucleus forward scattering amplitude
\begin{eqnarray}
\imag {\cal A}_{hA}&=&\sum_{n=1}^{A} \Bigg\lbrace
\frac{A!}{(A-n)!}(4 M \nu)^{-n+1}(\imag f_{hh})^n
\nonumber\\
&&\qquad\times\int \de^2b
        \int_{-\infty}^{\infty} \de z_1
        \int_{z_1}^{\infty} \de z_2
        \cdots
        \int_{z_{n-1}}^{\infty} \de z_n
\nonumber\\
&&\qquad\times
        \rho_n(\vec{b};z_1\dots z_n)
         \exp\left(
         \imag \frac{z_1-z_n}{\lambda}
        \right) \Bigg\rbrace   .
\end{eqnarray}

With the assumption that the hadronic
forward amplitudes $f_{hh}$ are dominated
by their imaginary parts
(see~\cite{Weise74}), we can use the
optical theorem to replace
\begin{equation}
 \imag f_{hh}\approx-2M\nu\sigma_{hN}.
\end{equation}
We finally  obtain the following
expression for the
hadron-nucleus cross section:
\begin{eqnarray} \label{eq:mulscat}
\sigma_{hA}&=&\sum_{n=1}^{A} \Bigg\lbrace \frac{A!}{(A-n)!}
        \left(-\frac12\right)^{n-1}
        \left(\sigma_{hN}\right)^n
\nonumber\\
&&\qquad\times { {\rm Re}}\bigg[\int \de^2b
        \int_{-\infty}^{\infty} \de z_1
        \int_{z_1}^{\infty} \de z_2
        \cdots
        \int_{z_{n-1}}^{\infty} \de z_n
\nonumber\\
&&\qquad\times
        \rho_n(\vec{b};z_1\dots z_n)
        \exp\left(
         \imag \frac{z_1-z_n}{\lambda}
        \right) \bigg]\Bigg\rbrace \nonumber\\
&=&
A \sigma_{hN}
        \left(
                1+\sum_{n=2}^{A} (-1)^{n-1}C_n
                \left(\sigma_{hN}\right)^{n-1}
        \right),
\end{eqnarray}
where
\begin{eqnarray}
C_n & = & \frac{(A-1)!}{2^{n-1}(A-n)!} {{\rm Re}}\Bigg[\int \de^2b
        \int_{-\infty}^{\infty} \de z_1
        \int_{z_1}^{\infty} \de z_2
        \cdots
        \int_{z_{n-1}}^{\infty} \de z_n
\nonumber\\
&&\qquad\times
        \rho_n(\vec{b};z_1\dots z_n)
        \exp\left(
         \imag \frac{z_1-z_n}{\lambda}
        \right) \Bigg].
\end{eqnarray}
Note that the exponential factor in
Eq.~(\ref{eq:mulscat})
oscillates rapidly if the coherence length
$\lambda$ of the hadronic scatterer is
small. In that case all terms in the
series with $n>1$ approximately vanish and
one finds $\sigma_{hA}\approx A
\sigma_{hN}$. In the small-$x$ region,
however, $\lambda$ increases and higher
order terms contribute, leading to a
reduction of $\sigma_{hA}$.

Let us take a closer look at the multiple
scattering series~(\ref{eq:mulscat}).  For
$n\ll A$ we may neglect the recoil motion
of the $A-n$ noninteracting nucleons.  In
the absence of nuclear correlations the
$n$-particle density is then approximated
by:
\begin{equation} \label{eq:densapprox}
\rho_n(\vec{b};z_1\dots z_n)
\approx \frac1{A^n}\prod_{i=1}^{n}\rho(\vec{b},z_i),
\end{equation}
where $\rho$ is the nuclear one-body
density, normalized as $\int \de^3x
\rho(\vec{x}) = A$.  For light nuclei only
single and double scattering contributions
are of importance.  The above
approximation may already be applied for
$A \geq 6$.  Furthermore the validity of
(\ref{eq:densapprox}) improves with
increasing $A$, since the number of
rescatterings $n_{\rm eff}$ that add
significantly to $\sigma_{hA}$ grows at
most as the nuclear diameter,
i.e.~${n_{\rm eff}\sim A^{1/3}}$.

For illustration, consider the multiple
scattering series (\ref{eq:mulscat}) with
a Gaussian density $\rho$
of
radius
$\langle r^2 \rangle^{1/2} =
\sqrt{3/2}\; a A^{1/3}$:
\begin{equation}
\rho
(\vec{r})
=\frac1{\pi^{3/2}a^3}
 \exp \left( - \frac{\vec{r}^2}{a^2 A^{2/3}}
\right) .
\label{eq:gaussshape}
\end{equation}
We obtain:
\begin{equation}
\sigma_{hA}\approx
A\sigma_{hN}
\left[
1- A^{1/3} \frac{\sigma_{hN}}{8\pi a^2} \frac{A-1}{A}
\exp \left(
-\frac {a^2 A^{2/3}}{2 \lambda^2}
\right)
+\dots\right].\label{eq:adep}
\end{equation}
We observe that the double scattering
contribution adds to the single scattering
term a negative correction, the magnitude
of which grows as $A^{1/3}$.  Furthermore
we notice that the shadowing correction
decreases rapidly if the coherence length
of the scatterer becomes small,
$\lambda < \langle r^2\rangle^{1/2}$.

\subsection{Shadowing in intermediate-mass and heavy nuclei}
\label{sec:heavy}

In the previous section we have prepared the tools
to calculate
total hadron-nucleus cross sections $\sigma_{hA}$
{}from the
respective hadron-nucleon cross sections
$\sigma_{hN}$. We can now proceed to calculate the nuclear
structure function as given by Eq.~(\ref{eq:F2A}).

We will first discuss heavier nuclei, making use of the
approximation in Eq.~(\ref{eq:densapprox}),
i.e.~replacing the $n$-particle density $\rho_n$
by a product of one-body densities.
We use two `extreme' parametrizations for these
nuclear matter densities: a Gaussian shape as in
Eq.~(\ref{eq:gaussshape}) for small $A$
and a square well shape for heavier nuclei. In
both cases we fit the mean square radii of these
density distributions to empirical  nuclear
radii~\cite{VriJagVr87}. Note that in earlier
calculations~\cite{PilWei90}
we have  used realistic densities and
included two-nucleon correlations,
but we found the resulting corrections in
both cases to be systematically very small.
With this as an input, we can now calculate the
shadowing ratios
\begin{equation}
R(x,Q^2)=
\frac{F_2^A(x,Q^2)}
{A\, F_2^N(x,Q^2)}.
\label{eq:shadratio}
\end{equation}
Figures~\ref{fig:He.Li.C.Ca}
and~\ref{fig:Xe}
show our calculated ratios for various
nuclei, together with experimental results
obtained by the NMC at
CERN~\cite{NMC91,Paic94} and the \mbox{E-665}
collaboration at
FNAL~\cite{E66592prl,E66592plb,Carroll93}
who have performed muon scattering measurements
focusing
on the small-$x$ region.
We see that for $x<0.1$ the ratio (\ref{eq:shadratio})
is generally
below one, i.e.\ shadowing occurs. In this
$x$~range we can apply the physical picture
introduced in Section~\ref{sec:labframe}:
The virtual photon interacts with the target
nucleus through hadronic fluctuations.
For small  $x$ the coherence length $\lambda$ of
the fluctuations becomes large enough to make
multiple scattering processes
contribute significantly.

However the shadowing ratio (\ref{eq:shadratio})
is not just a function of $x$ but also
depends (weakly) on $Q^2$.
We recall from our previous discussion that
the value of $Q^2$ basically selects that part
of the hadron mass spectrum which dominates the
interaction, and hence  determines which
cross sections $\sigma_{hN}$ contribute significantly
to the multiple scattering series.
{}From Section~\ref{sec:nucleon}
we note that $\sigma_{hN}$ depends not only on the
mass $\mu$ of the $q\bar q$ pair, but also on the
distribution of momenta within that pair. While
the averaged interaction cross sections decrease
as $\log(\mu^2)/\mu^2$  with increasing mass,
pairs which are asymmetric in the $q\bar q$ phase space
interact with large cross sections, even for large
$\mu$, and therefore produce strong shadowing.
This is the reason for the very weak overall $Q^2$
dependence of the shadowing effect.

The relevant experiments all operate on fixed
targets within a limited range of
muon energies, hence $Q^2$ is not
an independent parameter but depends
on the $x$-range considered.
We have taken this dependence into account by
inserting into our calculation the mean $Q^2$
values reported for the different
$x$-bins of the experiments.
With decreasing $x$ the accessible values for $Q^2$ also become small
(e.g.\ at $x = 0.005$, $Q^2 \sim 1 \GeV^2$ for the NMC experiment
from ref.~\cite{NMC91}).
Therefore the contributions of the low mass vector mesons
$\rho$, $\omega$ and $\phi$ dominate the observed shadowing
at $x<0.01$ as indicated in Figure 4 and 5.

The
NMC~\cite{NMC91,Paic94}
has analyzed
the  $Q^2$ dependence of shadowing by performing  linear
fits $R(x,Q^2)\approx a+b\,\ln Q^2$ to the data
for every $x$-bin. Fig.~\ref{fig:Qdep.He.C.Ca}
shows the slopes $b$ so obtained  in
comparison with our calculations.
We see that  the NMC data  are compatible with basically
no $Q^2$ dependence. Our
calculations give a very small positive slope,
i.e.\ a slow decrease in shadowing
with increasing $Q^2$ which is within
the range of the NMC data.

An NMC analysis of the structure function ratio
${\rm Sn}/{\rm C}$ is underway. It
combines data taken at several different
muon energies and provides considerably
better statistics.
Figure~\ref{fig:tin} shows
our predictions for this ratio in
about the kinematic region to be
covered.

Both the E-665 data on Xenon~\cite{E66592prl}
and
the recent
NMC data for Carbon and
Lithium~\cite{Paic94} extend to rather small
values of $x$ ($x<10^{-3}$). In this region a
saturation of the shadowing effect  becomes
apparent, with the ratio eventually approaching
the `photon point' i.e.\ the value observed in
the scattering of real photons on nuclei (see
Section~\ref{sec:photo} below).
Figures~\ref{fig:qsqdep}~a) and b) display the
shadowing ratio for Xenon,  computed at
various fixed values of the energy transfer
$\nu$, as a function of $x$ and a function
of $Q^2$, respectively. While the onset of
shadowing is controlled by the coherence
length $\lambda$, which enters as a
function of $x$, one sees that the
relevant variable for the saturation is $Q^2$.
As we have already argued, variation of
$Q^2$ basically scans the hadronic mass
spectrum of the photon.  Due to the
experimental constraints mentioned above,
small $x$ in practice always implies small
$Q^2$.
Saturation occurs at values of $Q^2$ less
than $m_{\rho}^2$, where the interaction
is dominated by multiple scattering of the
$\rho$~meson.

Here a remark is in order about contributions from inelastic
intermediate states to the multiple scattering series
(which we have neglected).
These are significant only at very small $x<10^{-3}$
and turn out to be small for heavy nuclei.
Their major contribution to shadowing increases
logarithmically with the energy transfer $\nu$ but  is
independent of $Q^2$ at small $Q^2<1\GeV^2$
(see Section~4.3 and ref.~\cite{NikZak91plb}).
Although the saturation value of $R(x,Q^2)$ at $x \ll 0.1$ may
therefore depend on $\nu$, the onset of the saturation is still
controlled by $Q^2$.

Figure~\ref{fig:adep.panel} shows the shadowing
ratio $12 F_2^A/ A F_2^{\rm C}$ for different nuclei
plotted against $\log A$ at several values
of $x$, together with preliminary NMC data \cite{SeiWit93}.
One sees that the dependence on $A$ is much
weaker than the behavior~${\sim A^{1/3}}$
one would derive by only considering the double
scattering term in~(\ref{eq:adep}).
In fact for heavier nuclei, higher order
contributions in the multiple scattering
series become important and partly
cancel  the effect of the double scattering
term.
The result is a much less pronounced $A$ dependence
that can be fitted by the expression
\begin{equation}
12 F_2^A/ A F_2^{{\rm C}} \approx a_x+b_x \ln A\,.
\label{eq:adep.linear}
\end{equation}
Figure~\ref{fig:adep.slopes} displays
the slopes $b_x$ in the different $x$ bins
resulting from our calculation and
those extracted in the preliminary NMC
analysis.

Note, however, that the
behavior according to Eq.~(\ref{eq:adep.linear})
cannot be correct in the limit of large $A$.
One should rather expect saturation of the
shadowing in this region, a hint of which can
be seen in our results.

On the whole, our model is able to reproduce
the shadowing phenomena in deep-inelastic scattering
on heavy nuclei
remarkably well. With respect to the $Q^2$
dependence, the pending release of the NMC data
on ${\rm Sn}$ may be interesting.

\subsection{Shadowing effects in deuterium}

Shadowing also occurs in deuterium, the most
weakly bound nucleus. Although small, this effect
is of special  interest since deep-inelastic
scattering from deuterium is used to determine
the structure function of the neutron.
With the assumption of isospin symmetry, the
proton and neutron structure functions together
reveal information on the quark flavor structure
of the nucleon.
These reasons and recent high precision
measurements of proton and deuteron structure
functions and their ratio \cite{NMC92,NMC94,Spentz94}
inspired a lively activity on this topic
(see e.g.
\cite{BadKwi92npb,
NNNZol92}).

Following our previous discussions, we now
calculate the deuteron structure function
$F_2^D$ at small values of $x$, taking
shadowing corrections explicitly into account.
Consequences for the experimental extraction of the
neutron structure function $F_2^n$ will then be
outlined briefly.

To calculate $F_2^D$  at $x<0.1$ we again need to know the
interaction cross section $\sigma_{hD}$ for the
scattering of a hadronic fluctuation from
the deuteron target (see Eq.~(\ref{eq:F2A})).
In addition to incoherent scattering from the two
nucleons,
$\sigma_{hD}$
includes  a coherent double scattering correction:
\begin{equation} \label{eq:sighD}
\sigma_{hD} = 2\,\sigma_{hN} - \delta \sigma_{hD}.
\end{equation}
{}From the multiple scattering series in
Eqs.~(\ref{eq:mulscat:zord},\ref{eq:mulscat:V(n)}) we find
\begin{equation} \label{eq:doublescatt}
\delta \sigma_{hD} =
    \frac{1}{2} \sum_{X} \frac{|f_{hX}|^2}{(2M\nu)^2}
    \,F_L\Bigl(1/\lambda (m_X)\Bigr).
\end{equation}
In contrast to
our treatment of multiple
scattering corrections in  heavy
nuclei in Section~\ref{sec:heavy}, we now take
inelastic intermediate states explicitly into
account.
The transition amplitude $f_{hX}$ describes
the  interaction of the hadronic state $h$
with a nucleon by which $h$ is converted into a
state $X$ with mass $m_X$.
The coherence length $\lambda(m_X)$ of the
hadronic state $X$ is defined as in
Eq.~(\ref{eq:coherence}); we explicitly
note its dependence on the mass of the propagating
state.
This coherence length
enters via the longitudinal form factor $F_L$
of the deuteron,
which can be written in terms of the deuteron
wave function as follows:
\begin{eqnarray} \label{eq:long.formfactor}
F_L(1/\lambda) &=&
\int \de z\,
\left|
        \psi(\vec{0}_{\perp}, z)
\right|^2
\exp\left(i\frac{z}{\lambda}\right),\\
\mbox{with}\quad \psi(\vec r) =
\psi_{J=1,M_J}(\vec{r}) &=&
    \frac 1{\sqrt{4\pi}}
    \left \{
        \frac{u(r)}{r}
        + \frac{w(r)}{r} \frac1{\sqrt 8} S_{12}({\hat r})
    \right \}
    \chi_{1,M_J}.
\end{eqnarray}
Here $u(r)$ and $w(r)$ are the $S$- and
$D$-wave
components of the deuteron wave function and
$\chi_{1,M_J}$ is its spin part.
$
S_{12}(\hat r )=
3
(\vec{\sigma}_1\cdot\hat r) \;
(\vec{\sigma}_2\cdot\hat r)
-(\vec{\sigma}_1\cdot\vec{\sigma}_2)
$ with $\hat r = \vec r/|\vec r|$ is the tensor
operator.
After taking the average over the target spin we find
\begin{equation} \label{eq:long.FF}
F_L(1/\lambda) =
\frac{1}{2\pi}
\int_0^{\infty} \frac{\de z}{z^2}\,
        \left[
                u^2(z) + w^2(z)
        \right] \cos\left(\frac{z}{\lambda}\right).
\end{equation}
Our expression for the double scattering
correction (\ref{eq:doublescatt}) can
be split into an elastic ($X=h$) and an
inelastic contribution ($X\ne h$). As in
Section~\ref{sec:glauber}, we assume that the
amplitudes $f_{hX}$ are strongly peaked in
forward direction and  dominated by their
imaginary parts.
We may then identify the inelastic contribution
with the cross section for inelastic diffractive
dissociation in the forward direction,
$h+N \to X+N$, and obtain:
\begin{equation} \label{eq:doublescatt2}
\delta \sigma_{hD} = \frac{1}{2} \sigma_{hN}^2
F_L\Bigl(1/\lambda(m_h=\mu)\Bigr)
+ \left. 8\pi \int \de m_X^2\,
  \frac{\de^2\sigma^{{\rm inel.}}
        _{h\rightarrow X}}
  {\de m_X^2 dt} \right|_{t=0}
\,F_L\Bigl(1/\lambda(m_X)\Bigr),
\end{equation}
where $t$ is the squared momentum transfer.

A well known feature of diffractive dissociation of hadrons and
photons is the $1/m_X^2$ mass spectrum at large $m_X$
(see e.g. \cite{Goulia83}). For hadron $h$
this reads
\begin{equation} \label{eq:3P}
\left.\frac{1}{\sigma_{h N}}
\frac
{\de^2\sigma^{{\rm inel.}}_{h\rightarrow X}}
{\de m_X^2 dt}\right|_{t=0} \approx
\frac{C}{m_X^2}\quad \mbox{for} \;m_X^2\gg \mu^2,
\end{equation}
where the constant
$C \approx 0.1\GeV^{-2}$
can be extracted
from the experimental analysis in
refs.~\cite{CoGoea81}.
We will use Eq.~(\ref{eq:3P}) to estimate the
inelastic contributions to the double scattering
correction in Eq.~(\ref{eq:doublescatt2}),
assuming that the $1/m_X^2$ behavior of the
diffractive cross section sets in at $m_X^2 =
(\mu + \Lambda)^2$.
In high energy hadron-nucleon scattering
experiments \cite{CoGoea81} one finds $\Lambda$
typically
to be of the order of $\Lambda \sim 1\GeV$.

With Eqs.~(\ref{eq:F2A},\ref{eq:sighD}) we may
now calculate the deuteron structure function
$F_2^D$.
We use a sample of different
deuteron wave functions
for this purpose:
those obtained
from the realistic Paris~\cite{LaLoRi80} and
Bonn~\cite{MaHoEl87} nucleon-nucleon potentials,
but also -- just for comparison -- the simple but
unrealistic Hulth\'en ansatz~\cite{HulSug57}.
We discuss our results for $F_2^D$
as above in terms of the structure function ratio
\begin{equation}
R^D(x,Q^2)=\frac{F_2^D(x,Q^2)}{2F_2^N(x,Q^2)}
=1 - \frac{\delta F_2^D(x,Q^2)}{2 F_2^N(x,Q^2)}.
\end{equation}

In Fig.~\ref{fig:deut.ratio.1}
we display $R^D(x,Q^2)$ as a function of $x$ for
different values of the momentum transfer $Q^2$.
We observe that $R^D(x,Q^2)<1$ in the range $x<0.1$,
i.e.\ the characteristic shadowing behavior.
The magnitude
of the effect is small but  depends on the
deuteron wave function used as an input.
For example at $x=0.01$ and $Q^2 = 4\GeV^2$ the
calculated shadowing effect varies between
$(1\mbox{--}2)\%$
(it amounts to $4\%$ for the naive Hulth\'en function).

This sensitivity is a consequence of significant
differences, for different potentials,
in the short distance behavior of the
deuteron density $\rho(r)= (u^2(r) + w^2(r))/(4\pi r^2)$, which
determines  the longitudinal form factor $F_L$ in
Eq.~(\ref{eq:long.FF}).
In Fig.~\ref{fig:deut.dens} we present $\rho(r)$
for the various deuteron wave functions, with
the densities differing  considerably for
$r<1\fm$. On the other hand we note that the
region $r<1\fm$ strongly influences  $F_L$ for
$\lambda>2\fm$ as can be seen from
Eq.~(\ref{eq:long.FF}).
As we have learned in~\ref{sec:glauber},
such  values of the propagation length control the
nuclear shadowing effect.

In Fig.~\ref{fig:deut.ratio.2}
we compare the calculated shadowing
effect with and without contributions from
inelastic intermediate states
at fixed $Q^2 = 4 \GeV^2$ for the Paris wave function.
We observe that inelastic states are important only
for $x<5\times 10^{-3}$.
For example at $x=10^{-3}$ they account for
about $\sim 20\%$ of the total shadowing effect.
Their contribution
decreases logarithmically with increasing
$x$ or, equivalently, with decreasing photon
energy $\nu$.

We may now briefly justify our
statement in section~\ref{sec:glauber}
that contributions
{}from inelastic intermediate states in multiple
scattering are small in  heavier nuclei.
In analogy to Eq.~(\ref{eq:doublescatt}) these
contributions are proportional to the
longitudinal nuclear form factor of the nucleus
which for the example of a Gaussian density reads~\cite{NikZak91plb}
\begin{equation}
F_L \Bigl(1/\lambda (m_X)\Bigr) \sim
\exp\left(-{\frac{\langle r^2\rangle}
  {3 \lambda(m_X)^2}}\right).
\end{equation}
{}From  Eq.~(\ref{eq:coherence}) we know that the
coherence lengths of intermediate states
decrease with the invariant mass of the
propagating states. Since the invariant mass of
the hadronic projectile $h$ is always smaller
than the mass of the diffractively  excited
inelastic intermediate states, the elastic
contribution  will naturally dominate the
multiple scattering process.
This dominance is  more pronounced
as the radius $\langle r^2 \rangle^{1/2}$ of the
nuclear target increases.

What are the consequences of the shadowing effect in
deuterium?
As already mentioned, the neutron structure function
$F_2^n$ is usually extracted from a comparison of the
deuteron and the proton structure function. Such an
analysis has been  performed recently by the NMC
experiment~\cite{NMC94} which
investigated the kinematic region $x<0.1$ with high
accuracy.
In this  analysis, however,
effects from nuclear shadowing in
deuterium have been ignored; the difference of proton and neutron
structure function was obtained by simply
taking
\begin{equation}
F_{2,\,{\rm NMC}}^{p-n}
\equiv (F_2^p - F_2^n)_{{\rm NMC}} =
2F_2^p - F_2^D.
\end{equation}
However, for small $x$
shadowing  must be taken into account. The
true structure function should therefore read:
\begin{equation}
F_{2}^{p-n} \equiv  F_2^p - F_2^n
= 2F_2^p - (F_2^D
+
\delta F_2^D)
= F_{2,\,{\rm NMC}}^{p-n} - \delta F_2^D.
\end{equation}
The shadowing correction $\delta F_2^D$
reduces the result with respect to
the quoted difference $F_{2,\,{\rm NMC}}^{p-n}$ (and
correspondingly the true neutron structure function
should be larger than the value obtained by the NMC).

The full symbols in Fig.~\ref{fig:Gottfried.shadcorr}
display the original NMC data for the difference
$F_{2,\,{\rm NMC}}^{p-n}$ as well as corrected results
with the shadowing term subtracted
for small $x$.
In our calculation we used $Q^2 = 4\GeV^2$,
as the NMC analysis operates with structure functions
interpolated to this value, and again a set of different
deuteron wave functions.
We notice that for small $x$ the structure
function difference
becomes small, so that the relative
size of the correction is of the order of $100\%$.
This is in good agreement with the expectation
that any deviation of $F_{2,\,{\rm NMC}}^{p-n}$
{}from zero in this $x$ region should be mostly due to the
shadowing effect.

Let us next consider the integral over the difference
of proton and neutron structure functions:
\begin{equation} \label{eq:IG0}
I_G(x,1) = \int_x^1 \frac{\de x'}{x'}
\left(
        F_2^p(x') - F_2^n(x')
\right).
\end{equation}
In the parton model $I_G$ can be rewritten in terms of
quark distributions:
\begin{equation} \label{eq:IG1}
I_G(x,1) = \frac{1}{3}\int_x^1 \de x'
\left( u_v(x')-d_v(x') \right)
+ \frac{2}{3}\int_x^1 \de x' \left(\bar u
(x')-\bar d(x')\right).
\end{equation}
The up and down valence quark distributions
$u_v=u-\bar u$ and $d_v=d-\bar d$ in the proton are
given by the difference of the respective quark and
antiquark distributions. To obtain Eq.~(\ref{eq:IG1})
we have used isospin symmetry to relate proton and
neutron quark distributions.

If the first moments of the up and down sea quark
distributions are approximately equal -- this
is fulfilled trivially if one assumes the sea
to be SU(2)-flavor symmetric -- the second term in
Eq.~(\ref{eq:IG1})  vanishes for $x\to 0$,
and one arrives at the Gottfried sum
rule~\cite{Gottfr67}:

\begin{equation}
I_G = I_G(0,1) = \frac{1}{3}\int_0^1 \de x'
\left(u_v(x')-d_v(x')\right)
         = \frac{1}{3}.
\end{equation}
Without taking shadowing into account, the NMC found
$I_G^{\rm exp.}(0,1)=0.235\pm 0.026$~\cite{NMC94}.
This includes contributions from the
unmeasured regions $x > 0.8$ and $x<0.004$.
A smooth extrapolation of $F_2^n/F_2^p$  for $x\rightarrow 1$
yields $I_G(0.8,1) = 0.001\pm 0.001$, while
within conventional Regge theory $I_G(0,0.004) = 0.013 \pm 0.005$
is found \cite{NMC94}.

The  deviation of the  Gottfried sum
$I_G^{\rm exp.}$ from the na\"\i{}ve expectation $1/3$
has been the target of some activity,
see e.g.~\cite{EllSti91}.

Let us now consider the impact of our
corrections for deuteron shadowing on  the extraction
of the Gottfried sum.
As discussed above, for $x<0.1$ they
significantly reduce the difference $F_2^{p-n}$
with respect to the values  used by the NMC collaboration.

\sloppy
The open symbols in Fig.~\ref{fig:Gottfried.shadcorr}
show the NMC results for the integrals $I_G(x,0.8)$
together with our corrected values.
In Tab.~\ref{tab:gottfried}
we give  the corrections $\Delta
I_G(0.004,0.1)$ due to shadowing
obtained for different deuteron wave functions.
They reduce the extracted value for the Gottfried sum
by $10\%$ or more.
This  sizable correction is due to the fact that the
structure function difference in the integral
in (\ref{eq:IG0}) is weighted by a factor
$1/x$. We see that the shadowing corrections further  enhance
the deviation of the Gottfried sum
{}from the na{\"\i}ve value $I_G = 1/3$.

\fussy

The E665 collaboration has also measured
the structure function ratio
$F_2^d/F_2^p$, for which preliminary data
are now available~\cite{Spentz94}. Their $x$-range extends down
to $10^{-6}$, with the average $Q^2$ strongly
dependent on $x$
($\langle Q^2\rangle=0.002\GeV^2$ for the
lowest $x$ bin $10^{-6}<x<10^{-5}$).
Figure~\ref{fig:deut.E665} shows the ratio
$F_2^d/2F_2^p$ as obtained  by the E665 group.
The small-$x$ behavior of these data is in excellent
agreement with the results of our model calculation.
In comparing with Figs.~\ref{fig:deut.ratio.1}
and~\ref{fig:deut.ratio.2} we note that the logarithmic
growth of shadowing for $x\to 0$
displayed there is a result of
plotting the ratio at constant $Q^2$, which implies
a photon energy $\sim 1/x$.
In this  experiment the muon (and hence the
photon)  energy  is  limited
and therefore the contributions
{}from  inelastic intermediate states
are bounded.

\section{Nuclear shadowing of real photons at high energy}
\label{sec:photo}
In deep-inelastic scattering, the lepton
beam is a source of highly energetic {\em
  virtual\/} photons.  We have
conveniently written the structure
function in terms of a total cross section
for the interaction of the virtual photon
with the target (see
Eq.~\ref{eq:sigmagammaN1}).  In the limit
$Q^2\to0$ this cross section becomes
physical, describing real photoproduction
processes.

While hadronic vacuum fluctuations
decouple from the photon in the limit
$Q^2\rightarrow 0$, so as to keep the real
photon massless, shadowing still occurs
via the production of quark-antiquark
pairs on nucleons and their subsequent
propagation through the nucleus. The
coherence length of such hadronic states
of mass $\mu$, produced by a real photon
of energy $\nu$, is
\begin{equation}
\lambda = \frac{2 \nu}{\mu^2}.
\end{equation}
This becomes largest for those states with
the smallest mass. For the $\rho$ meson
with $\mu^2=m_{\rho}^2$ we find that
$\lambda$ reaches typical internucleon
distances for $\nu > 3\GeV$.  We therefore
expect multiple scattering to reduce the
cross section for nuclear photoproduction
$\sigma_{\gamma A}$ with respect to $A$
times the photoproduction cross section on
free nucleons $\sigma_{\gamma N}$.

Let us discuss this more quantitatively.
The expression for $\sigma_{\gamma N}$ at
large $\nu$ is obtained from
Eq.~(\ref{F2N}) taking the limit $Q^2 \to
0$. This yields
\begin{eqnarray}
  \sigma_{\gamma N} &=&
  \Bigl(\sigma_{\gamma
    N}\Bigr)_{\rm VDM} +
  \Bigl(\sigma_{\gamma
    N}\Bigr)_{\rm cont.} \nonumber \\
  &=& \sum_V
  \frac{4\pi\alpha_{\rm em}}{g_V^2}
  \sigma_{VN} +4\pi\alpha_{\rm em}
  \int_{\mu_0^2}^{\infty}\de\mu^2 \int_0^1
  \de\alpha \frac{\Pi^{\rm cont.}(\mu^2)}{\mu^2}
  \sigma_{hN}(\mu^2,\alpha).
\label{eq:sigmaGammaN}
\end{eqnarray}
The VMD term ($\sim 90 {\rm \mu b}$) is
about twice as large as the continuum
contribution.

As in the case of deep-inelastic
scattering, we can calculate the cross
section for photoproduction on nuclei by
replacing the hadron-nucleon cross
sections in Eq.~(\ref{eq:sigmaGammaN}) by
the corresponding hadron-nucleus cross
sections, obtained via the multiple
scattering formalism as outlined
in~\ref{sec:glauber}.  To discuss nuclear
effects it is common to consider the ratio
\begin{equation}
  \frac{A_{\rm eff}}{A} =
  \frac{\sigma_{\gamma
      A}}{A\,\sigma_{\gamma N}}
\end{equation}
In our calculation of $\sigma_{\gamma N}$
we use parameters which were fixed through
our fit to the nucleon structure function
$F_2^N$ in Section~\ref{sec:nucleon}. To
calculate $\sigma_{\gamma A}$ we employ
similar nuclear densities as in the case
of nuclear deep-inelastic scattering in
Sec.~\ref{sec:heavy}.

The $A$-dependence of photoproduction on nuclei
has been measured by several
groups~\cite{HeMeNa71}.  Figure~\ref{fig:photo} displays our
results of the shadowing ratio
$A_{\rm eff}/A$ calculated for C,
Cu and Pb target nuclei, together with
various experimental data.  We observe
significant shadowing for $\nu>2\GeV$.
The effect is well described within our
model.  Its results are quite similar to
those of earlier VMD
calculations~\cite{BaSpYe78}, which should
not be much of a surprise, due to the
dominant r\^ole of the VMD term
noted
above.

Shadowing grows stronger for higher photon
energies $\nu$, eventually approaching
some saturation value, apart from
logarithmic corrections due to inelastic
contributions to multiple scattering (see
discussion in Sections 4.2 and 4.3).  This
is due to the fact that the coherence
length $\lambda$ governing multiple
scattering processes is now directly
proportional to $\nu$.

The validity of the picture developed here is restricted to
$\nu > 2 \GeV$. For smaller photon energies, photonuclear dynamics
is governed by the excitation and propagation of nucleon resonances
in nuclei.

\section{Summary}

Shadowing at small values $x<0.1$ of the
Bjorken variable is the most prominent
nuclear effect seen in deep-inelastic
lepton scattering from nuclear targets.
We have developed a phenomenology of
nuclear shadowing, expressed in the
laboratory frame, which makes use of the
full hadronic spectrum of virtual photons.
Our framework unifies the vector meson
dominance picture with the concept of
color transparency applied to the
quark-antiquark continuum part of the
photon spectral function.

\bigskip
\noindent
Our results are summarized as follows:
\begin{enumerate}
\item[i)] In our lab frame approach
  shadowing arises from the coherent
  multiple scattering of quark-antiquark
  fluctuations through the nuclear target.
  A satisfactory description of nuclear
  structure functions at small $x$ is
  achieved for a large variety of nuclei,
  both light and heavy.

\item[ii)] Vector mesons dominate at small
  $Q^2 \lsim 1 \GeV^2$.  At large $Q^2$
  the quark-antiquark continuum becomes
  important.  The combination of both
  resonant and continuum parts of the
  hadronic photon spectrum is crucial in
  order to obtain the almost negligible
  overall $Q^2$ dependence of the
  shadowing effect, whereas vector meson
  dominance alone would imply decreasing
  shadowing with increasing $Q^2$.

\item[iii)] Contributions to shadowing
  {}from inelastic intermediate states in
  the multiple scattering chain give only
  small corrections for heavy nuclei, but
  they need to be taken into account for
  light nuclei, in particular for the
  deuteron.

\item[iv)] Consistency is found with the
  observed shadowing for interactions of
  real photons with nuclei at high
  energies.

\item[v)] Special emphasis has been
  directed to shadowing effects in
  deuterium. In the extraction of the
  neutron structure function from deuteron
  data at small $x$, such effects must be
  taken into account carefully. We find
  strong sensitivity to the short distance
  behavior of the deuteron wave function;
  the use of realistic nucleon-nucleon
  potentials is therefore absolutely
  necessary for a reliable description.
  Shadowing effects of $1\mbox{--}2\%$
  imply corrections of the order of $10\%$
  in the Gottfried sum. This correction
  further increases the already
  established discrepancy with the naive
  parton model.

\end{enumerate}

\noindent We would like to thank S.~Kulagin,
W.~Melnitchouk, N.~N.~Nikolaev and
A.~W.~Thomas for helpful discussions and
comments.


\clearpage\section*{Tables}
\begin{table}[h]
\centering\leavevmode
\smallskip
\begin{tabular}{cr@{$\pm$}lrlrl}
\hline\hline
$V$
& \multicolumn{2}{c}{$m_V/\MeV$\cite{PDG94}}
& \multicolumn{1}{c}{$g_V^2/4\pi$}
& \multicolumn{1}{c}{$\sigma_{VN}/\mb$}
\\
\hline
$\rho$   &  769.9   & 0.8    &  2.0
              & \quad $22\mbox{--}27$ & \cite{BaSpYe78}\\
$\omega$ &  781.9  & 0.1  & 23.1
              & \quad $25\mbox{--}27 $ & \cite{BaSpYe78}\\
$\phi$   & 1019.41 & 0.01 & 13.2
              & \quad $9\mbox{--}12$ & \cite{BaSpYe78}\\
$J/\psi$ & 3096.88  & 0.04 & 10.5
              & \quad $2.2\pm0.7$ & \cite{EMC83ccbar}\\
$\psi'$  & 3686.0   & 0.1   & 30.6
              & \quad $\sim 1.3$ & \cite{HoLeWi85}\\
\hline\hline
\end{tabular}
\caption{%
Vector meson properties: masses, couplings to the photon,
and total vector meson-nucleon cross sections. The coupling constants
$g_V$ are derived from the $V\rightarrow
e^+ e^-$ decay widths \protect
\cite{PDG94}.}
\label{tab:vmesons}
\end{table}
\begin{table}[h]
\centering\leavevmode
\smallskip
\begin{tabular}{c@{\qquad}ccc}
\hline\hline
&\multicolumn{3}{c}{$\Delta I_G(0.004,0.1)$}
\\
$I_G^{{\rm exp.}}(0.004,0.8)$ \cite{NMC94}&
Bonn$(1)$&
Bonn$(2)$&
Paris\\
\hline
$0.221\pm 0.021$ &
$-0.022$ & $-0.039$ & $-0.017$ \\
\hline\hline
\end{tabular}
\caption{Shadowing corrections $\Delta
  I_G$ for the Gottfried
sum obtained for
various deuteron wave functions compared to the experimental value
$I_G^{{\rm exp.}}$. Here ``Bonn$(1)$''
refers to the one-boson
exchange
Bonn potential, ``Bonn$(2)$'' is the full potential including
explicit two-pion exchange etc. \protect\cite{MaHoEl87}.
The Paris potential is taken from ref.~\protect\cite{LaLoRi80}.
}
\label{tab:gottfried}
\end{table}
\clearpage\section*{Figure captions}

\begin{enumerate}
\item The two possible time orderings for
  the interaction of a (virtual) photon
  with a nucleon or nuclear target: (a)
  the photon hits a quark in the target,
  (b) the photon creates a $q\bar q$ pair
  that subsequently interacts with the
  target.
\label{fig:timeorder}
\item Nucleon structure function for small
  $x$ plotted against $Q^2$. The solid
  line is the full result of our
  calculation. The contribution of vector
  mesons is indicated by the dashed line.
  We compare to NMC data from
  ref.~\cite{NMC92}.
\label{fig:fitnucleon}
\item Contribution ${{\cal A}_h^{(n)}}$ to
  the multiple scattering series: the
  hadronic projectile scatters from $n$
  nucleons inside the target nucleus.
\label{fig:ms1}
\item Our results for shadowing in He, Li,
  C, and Ca compared to available
  experimental data
  \cite{NMC91,Paic94,Carroll93}.  The
  dashed curves show the shadowing caused
  by the vector mesons $\rho$, $\omega$
  and $\phi$ only.
\label{fig:He.Li.C.Ca}
\item Shadowing in Xenon. Data are from
  the FNAL E-665 experiment
  \cite{E66592prl,E66592plb}.  The vector
  meson contribution is shown by the
  dashed curve.
\label{fig:Xe}
\item The slope $b=\de R / \de \ln Q^2$
  indicating the $Q^2$ dependence
  of the shadowing ratio for He, Li, C,
  and Ca
  extracted by the
  NMC~\protect\cite{NMC91,Paic94} for
  various $x$-bins together with our
  results.
\label{fig:Qdep.He.C.Ca}
\item $Q^2$ dependence of the shadowing
  ratio ${\rm Sn}/{\rm C}$ as predicted by
  our model, in the region to be covered
  by recent NMC data.
\label{fig:tin}
\item The shadowing ratio in Xe (a) as a
  function of $x$ at fixed $\nu$; (b) as a
  function of $Q^2$ at fixed $\nu$.
\label{fig:qsqdep}
\item The shadowing ratio as a function of
  the nuclear mass number $A$ for several
  $x$-bins. Experimental data are
  preliminary NMC results from
  \protect\cite{SeiWit93}.
\label{fig:adep.panel}
\item Our results for the slopes $b_x$ of
  the $A$ dependence relative to Carbon
  obtained from
  Eq.~(\protect\ref{eq:adep.linear})
  compared to the NMC data in
  ref.~\protect\cite{SeiWit93}.
\label{fig:adep.slopes}
\item Shadowing ratios in deuterium
  plotted against $x$ at different values
  of $Q^2$ and for different deuteron wave
  functions. Here ``Bonn$(1)$'' refers to
  the one-boson exchange Bonn potential,
  ``Bonn$(2)$'' is the full potential
  including explicit two-pion exchange
  etc. \cite{MaHoEl87}.  The Paris
  potential is taken from
  ref.\cite{LaLoRi80}.
\label{fig:deut.ratio.1}
\item The density
  $\rho(r)=(u^2(r)+w^2(r))/4\pi r^2$
  corresponding to different
  parametrizations of the deuteron wave
  function.
\label{fig:deut.dens}
\item The shadowing ratio for deuterium at
  fixed $Q^2=4\GeV^2$ calculated for the
  Paris wave function.  The full line
  includes inelastic intermediate states.
  The dashed curve was obtained by taking
  only elastic intermediate states into
  account.
\label{fig:deut.ratio.2}
\item Gottfried sum and shadowing: the
  filled symbols display the structure
  function difference $F_2^{p-n}(x)$, with
  the circles representing the original
  NMC data~\protect\cite{NMC94}, squares
  and diamonds include shadowing
  correction using different deuteron wave
  functions.  Open symbols show the
  respective values for the integral
  $I_G(x,0.8)$.
\label{fig:Gottfried.shadcorr}
\item Our result for $R^D(x,Q^2)$ compared
  to recent FNAL E-665 \cite{Spentz94} and
  NMC data \cite{NMC94} for the ratio
  $F_2^d/2F_2^p$.
\label{fig:deut.E665}
\item The shadowing ratio for the
  absorption of real photons on nuclei as
  calculated in our model and measured by
  various
  groups~\protect\cite{HeMeNa71}
  plotted against the photon energy $\nu$.
\label{fig:photo}
\end{enumerate}


\clearpage

\begin{thebibliography}{10}

\bibitem{NMC91}
{NMC, P. Amaudruz} et~al.,
\newblock Z. Phys. C 51 (1991) 387;\\
\newblock {NMC, P. Amaudruz} et~al.,
\newblock A re-evaluation of the nuclear structure function ratios
for D, He, ${}^6$Li, C and Ca,
\newblock {\em submitted to Nucl.\ Phys~B.}

\bibitem{NMC92}
{NMC, P. Amaudruz} et~al.,
\newblock Phys. Lett. {B} 295 (1992) 195.

\bibitem{NMC94}
{NMC, M. Arneodo} et~al.,
\newblock Phys. Rev. {D} 50 (1994) R1.

\bibitem{Paic94}
A.~Pai{\'c},
\newblock Structure function ratios
{$F_2^{\rm Li}/F_2^{\rm D}$} and
{$F_2^{\rm C}/F_2^{\rm D}$} at low $x_{\rm bj}$,
\newblock in {\em {PAN XIII}, Particles
  and Nuclei, {XIII} international
  conference},
edited by A.~Pascolini, Singapore, 1994, World Scientific;\\
\newblock {NMC, M. Arneodo} et~al.,
\newblock The structure function ratios
{$F_2^{\rm Li}/F_2^{\rm D}$} and
{$F_2^{\rm C}/F_2^{\rm D}$} at small $x$,
\newblock {\em submitted to Nucl.\ Phys.~B.}

\bibitem{E66592prl}
{FNAL E665, M. R. Adams} et~al.,
\newblock Phys. Rev. Lett. 68 (1992) 3266.

\bibitem{E66592plb}
{FNAL E665, M. R. Adams} et~al.,
\newblock Phys. Lett. {B} 287 (1992) 375.

\bibitem{Carroll93}
{FNAL E665, Timothy J. Carroll},
\newblock Ratios of cross sections of
carbon,
calcium and lead at low {$x_{\rm
  Bj}$} in inelastic muon scattering,
\newblock FERMILAB-Conf-93/166-E, 1993,
\newblock presented at the {\em 28th
  Rencontres de Moriond,
  QCD and High Energy
  Hadronic Interactions}, {Les Arcs, Savoie, France}.

\bibitem{Spentz94}
{FNAL E665, P. Spentzouris} et~al.,
\newblock Structure functions and
structure
function ratio {$F_2^n/F_2^p$} at
  small $x_{\rm bj}$ and ${Q}^2$ in muon-nucleon scattering,
\newblock FERMILAB-Conf-94/220-E, 1994,
\newblock presented at {\em Intersections of Particle and Nuclear
Physics,} St.
  Petersburg, FL.

\bibitem{BadKwi92rmp}
B.~Bade{\l}ek and J.~Kwieci\'{n}ski,
\newblock Rev. Mod. Phys. 64 (1992) 927;
\newblock
M.~Arneodo,
\newblock Phys. Rep. 240 (1994) 301.

\bibitem{MueQiu86}
A.~H. Mueller and J.~Qiu,
\newblock Nucl. Phys. {B} 268 (1986) 427;
\newblock
J.~Qiu,
\newblock Nucl. Phys. {B} 291 (1987) 746.

\bibitem{SakSch72}
J.~J. Sakurai and D.~Schildknecht,
\newblock Phys. Lett. {B} 40 (1972) 121;
\newblock
P.~Ditsas, B.~J. Read, and G.~Shaw,
\newblock Nucl. Phys. {B} 99 (1975) 85;
\newblock
C.~L. Bilchak, D.~Schildknecht, and J.~D. Stroughair,
\newblock Phys. Lett. {B} 214 (1988) 441,
\newblock Phys. Lett. {B} 233 (1989) 461.

\bibitem{Shaw89}
G.~Shaw,
\newblock Phys. Lett. {B} 228 (1989) 125,
\newblock Phys. Rev. {D} 47 (1993) 3676.

\bibitem{BaSpYe78}
T.~H. Bauer, R.~D. Spital, D.~R. Yennie, and F.~M. Pipkin,
\newblock Rev. Mod. Phys. 50 (1978) 261.

\bibitem{FraStr89}
L.~L. Frankfurt and M.~I. Strikman,
\newblock Nucl. Phys. {B} 316 (1989) 340.

\bibitem{BrodLu90}
S.~J. Brodsky and H.~J. Lu,
\newblock Phys. Rev. Lett. 64 (1990) 1342.

\bibitem{NikZak91zpc}
N.~N. Nikolaev and B.~G. Zakharov,
\newblock Z. Phys. C 49 (1991) 607.

\bibitem{BadKwi92npb}
B.~Bade{\l}ek and J.~Kwieci\'{n}ski,
\newblock Nucl. Phys. {B} 370 (1992) 278.

\bibitem{MelTho93}
W.~Melnitchouk and A.~W. Thomas,
\newblock Phys. Lett. {B} 317 (1993) 437.

\bibitem{KuPiWe94}
S.~A. Kulagin, G.~Piller, and W.~Weise,
\newblock Phys. Rev. {C} 50 (1994) 1154.

\bibitem{PilWei90}
G.~Piller and W.~Weise,
\newblock Phys. Rev. {C} 42 (1990) R1834,
\newblock Nucl. Phys. {A} 532 (1991) 271c.

\bibitem{DonSha78}
A.~Donnachie and G.~Shaw,
\newblock Generalized vector dominance,
\newblock in
{\em Electromagnetic Interactions of Hadrons, Vol. 2}, edited by
A.~Donnachie and G.~Shaw, page 169, Plenum Press, New York, 1978.

\bibitem{Nikola92}
N.~N. Nikolaev,
\newblock Comments on Nuclear and Particle Physics 21 (1992) 41;
\newblock
L.~L. Frankfurt, G.~A. Miller, and M.~Strikman,
\newblock Ann. Rev. of Nucl. and Part.
Sci. 44 (1994) 501.

%
\bibitem{DaBaea94}
S.~Dasu et al.,
\newblock Phys. Rev. {D} 59 (1994) 5641.

\bibitem{Gribov70}
V.~N. Gribov,
\newblock JETP 30 (1970) 709.

\bibitem{BjoKog73}
J.~D. Bjorken and J.~Kogut,
\newblock Phys. Rev. {D} 8 (1973) 1341.

\bibitem{PDG94}
{PDG, L. Montanet} et~al.,
\newblock Phys. Rev. {D} 50 (1994).

\bibitem{EMC83ccbar}
{EMC, J. J. Aubert} et~al.,
\newblock Nucl. Phys. {B} 213 (1983) 1.

\bibitem{HoLeWi85}
S.~D. Holmes, W.~Lee, and J.~E. Wiss,
\newblock Ann. Rev. of Nucl. and Part. Sci. 35 (1985) 397.

\bibitem{KoNeNi93}
B.~Z. Kopeliovich, J.~Nemchick, N.~N. Nikolaev, and B.~G. Zakharov,
\newblock Phys. Lett. {B} 309 (1993) 179.

\bibitem{DavNik78}
G.~V. Davidenko and N.~N. Nikolaev,
\newblock Nucl. Phys. {B} 135 (1978) 333.

\bibitem{Bertoc72}
L.~Bertocchi,
\newblock Nuovo Cim. 11A (1972) 45;
\newblock
J.~H. Weis,
\newblock Acta Physica Polonica B7 (1976) 851.

\bibitem{MuAyGu75}
P.~V.~R. Murthy, C.~A. Ayre, H.~R.
Gustafson, L.~W. Jones,
and M.~J. Longo,
\newblock Nucl. Phys. {B} 92 (1975) 269.

\bibitem{Nikola86}
N.~N. Nikolaev,
\newblock Z. Phys. C 32 (1986) 537.

\bibitem{Weise74}
W.~Weise,
\newblock Phys. Rep. 13 (1974) 53.

%
\bibitem{VriJagVr87}
H.~de~Vries, C.~W. de~Jager, and C.~de~Vries,
\newblock Atomic Data and Nuclear Data Tables 36 (1987) 495.

\bibitem{NikZak91plb}
N.~N. Nikolaev and B.~G. Zakharov,
\newblock Phys. Lett. {B} 260 (1991) 414.

\bibitem{SeiWit93}
{NMC, R. Seitz} and A.~Witzmann,
\newblock The ratio {$F_2^n/F_2^p$} and the $A$-dependence of nuclear
ratios measured by the NMC, 1993,
\newblock
Presented at the {\em 28th Rencontres de Moriond, QCD and High Energy
  Hadronic Interactions}, {Les Arcs, Savoie, France}.

\bibitem{NNNZol92}
N.~N. Nikolaev and V.~R. Zoller,
\newblock Z. Phys. C 56 (1992) 623;
\newblock
V.~R. Zoller,
\newblock Z. Phys. C 54 (1992) 425;
\newblock
W.~Melnitchouk and A.~W. Thomas,
\newblock Phys. Rev. {D} 47 (1993) 3783;
\newblock
V.~Barone, M.~Genovese, N.~N. Nikolaev, and E.~Predazzi,
\newblock Phys. Lett. {B} 321 (1994) 137;
\newblock
B.~Bade{\l}ek and J.~Kwieci{\'n}ski,
\newblock Phys. Rev. {D} 50 (1994) R4.

\bibitem{Goulia83}
K.~Goulianos,
\newblock Phys. Rep. 101 (1983) 171.

\bibitem{CoGoea81}
R.~L. Cool, K.~Goulianos, S.~L. Segler, H.~Sticker, and S.~N. White,
\newblock Phys. Rev. Lett. 47 (1981) 701;
\newblock
T.~J. Chapin et~al.,
\newblock Phys. Rev. {D} 31 (1985) 17.

\bibitem{LaLoRi80}
M.~Lacombe, B.~Loiseau, J.~M. Richard, and R.~V. Mau,
\newblock Phys. Rev. {C} 21 (1980) 861.

\bibitem{MaHoEl87}
R.~Machleidt, K.~Holinde, and C.~Elster,
\newblock Phys. Rep. 149 (1987) 1.

\bibitem{HulSug57}
L.~Hulth{\'e}n and M.~Sugawara,
\newblock The two nucleon problem,
\newblock in {\em Handbuch der Physik},
edited by S.~Fl{\"u}gge, volume~39,
  Springer, Berlin, 1957.

\bibitem{Gottfr67}
K.~Gottfried,
\newblock Phys. Rev. Lett. 18 (1967) 1174.

\bibitem{EllSti91}
S.~D. Ellis and W.~J. Stirling,
\newblock Phys. Lett. {B} 256 (1991) 258;
\newblock
E.~M. Henley and G.~A. Miller,
\newblock Phys. Lett. {B} 251 (1990) 453;
\newblock
W.-Y.~P. Hwang, J.~Speth, and G.~E. Brown,
\newblock Z. Phys. A 339 (1991) 383;
\newblock
S.~Kumano and J.~T. Londergan,
\newblock Phys. Rev. {D} 46 (1992) 457;
\newblock
J.~Levelt, P.~J. Mulders, and A.~W. Schreiber,
\newblock Phys. Lett. {B} 263 (1991) 498;
\newblock
B.-Q. Ma, A.~Sch{\"a}fer, and W.~Greiner,
\newblock Phys. Rev. {D} 47 (1993) 51;
\newblock
W.~Melnitchouk and A.~W. Thomas,
\newblock Phys. Rev. {D} 47 (1993) 3794;
\newblock
A.~Szczurek and J.~Speth,
\newblock Nucl. Phys. {A} 555 (1993) 249.

\bibitem{HeMeNa71}
V.~Heynen, H.~Meyer, B.~Naroska, and D.~Notz,
\newblock Phys. Lett. 34B (1971) 651;
\newblock
G.~R. Brookes et~al.,
\newblock Phys. Rev. {D} 8 (1973) 2826;
\newblock
S.~J. Michalowski et~al.,
\newblock Phys. Rev. Lett. 39 (1977) 737;
\newblock
D.~O. Caldwell et~al.,
\newblock Phys. Rev. {D} 7 (1973) 1362;
\newblock
D.~O. Caldwell et~al.,
\newblock Phys. Rev. Lett. 42 (1979) 553;
\newblock
E.~A. Arakelyan et~al.,
\newblock Phys. Lett. 79B (1978) 143;
\newblock
N.~Bianchi et~al.,
\newblock Phys. Lett. 325B (1994) 333.
\end{thebibliography}
\end{document}